# Chromium hardening and Peierls mechanism for basal slip in sapphire (a-Al2O3) at temperatures between 900 and 1500 °C

M. Castillo Rodríguez [a], A. Muñoz [a,*], J. Castaing [b], P. Veyssière [c], A. Domínguez Rodríguez [a]

a *Departamento Física de la Materia Condensada, Universidad de Sevilla, Apdo. 1065, 41080 Seville, Spain*

b *C2RMF, CNRS UMR 171, Palais du Louvre, 14 quai François Mitterrand 75001 Paris, France*

c *LEM, CNRS UMR 104, ONERA, BP 72, 92322 CHATILLON Cedex, France*

**Abstract**

The plastic deformation of $Cr^{3+}$-doped $a$-$Al_2O_3$ (ruby) with four distinct Cr concentrations has been studied at temperatures between 900 and 1500 °C. The mechanical tests indicate that the processes of dislocation multiplication and the adjustments of the slip velocity of these dislocations to the imposed strain rate in undoped $a$-$Al_2O_3$ are not significantly affected by the Cr concentration. The yield stress increment is approximately constant with temperature and the hardening increases with the Cr concentration. The critical resolved shear stress (CRSS) is fitted to a model based on dislocation glide controlled by the nucleation and propagation of kink pairs, modified to include the hardening due to Cr. Satisfactory fits are achieved with adjustable parameters close to those employed for undoped $a$-$Al_2O_3$.

**1. Introduction**

During the last decades, the plastic deformation of sapphire ($a$-$Al_2O_3$) has been studied by many authors because it is a crystalline material with low symmetry whose deformation can take place by slip in the basal plane as much as in the prismatic planes, and also by twinning along rhombohedral (pyramidal) planes. Pyramidal slip and basal twinning can also be activated in this crystal. $(0\,0\,0\,1)1/3\langle 2\bar{1}\bar{1}0\rangle$ basal slip is the primary deformation system for temperatures above 700 °C.[1–7] Prism plane slip along $\{1\,2\bar{1}\,0\}\langle 1\bar{0}1\,0\rangle$ is the easiest slip system below 700 °C.[3,4] The $\langle 1\bar{0}1\,0\rangle$ Burgers vector for prism plane slip, which is the fourth in length in sapphire, is dissociated into three partial dislocations with collinear $1/3\langle 1\bar{0}1\,0\rangle$ Burgers vector.[8] The Peierls mechanism has been frequently invoked to explain the movement of dislocations in sapphire, as much in the basal plane as in the prismatic plane, at intermediate temperatures.[3–6] Rhombohedral twinning, the softest deformation mode of sapphire above 500 °C, is activated under a shear stress of 10 MPa, independent of test temperature; this stress can, however, be considerably increased, by a factor of 5–10, in the presence of a high density of dislocations in the material.[9] Doping with small concentrations of elements of different atomic radii or different valence allows to create, in a controlled way, point defects acting as obstacles to dislocation glide. Solid solution hardening in sapphire ($a$-$Al_2O_3$) has been first observed by Wachtman et al.[10] in Cr-doped specimens deformed in basal slip. Klassen-Neklyudova et al.11 showed that the addition of Cr raises the yield point of specimens tested in tension between 1650 and 1900 ◦C. Radford et al.12 studied the influence of var- ious impurities (Fe, Ni, Cr, Ti and Mg) in compression between 1300 and 1600 ◦C. They concluded that hardening is related to either the difference in size of the ions isovalent to $Al^{3+}$, or to the valence of the solute in the case of aliovalent ions. Pletka et al.13 studied the hardening of sapphire doped with $Cr^{3+}$, $Ti^{3+}$ and $Ti^{4+}$, deformed by basal slip, between 1300 and 1500 ◦C for low Cr concentrations, between 1400 and 1500 ◦C for large Cr con- centrations, and above 1500 ◦C for Ti-doped sapphire. As for the isovalent $Cr^{3+}$ and $Ti^{3+}$ ions, it was concluded that the harden- ing is consistent with the models of Fleischer14 and Labusch,15 i.e. the hardening is due to the elastic strain field of the solute atom. Hardening by the aliovalent $Ti^{4+}$ is more effective due to the additional electrical interaction between dislocations and charged defects.

In the present work, we have studied the dependence of the yield stress of $Cr^{3+}$ containing sapphire (rubies) deformed in basal slip between 900 and 1500 ◦C. Four Cr concentrations have been investigated between 60 and 9540 mol ppm. In an attempt to investigate situations where the Peierls mechanism should take an important part, the temperature range has been expanded well below temperatures explored so far. In order to explain the temperature dependence of the critical resolved shear stress (CRSS) in sapphire, Mitchell et al.16–18 have developed a model based on that designed by Hirth and Lothe19 for dislocation glide by nucleation and propagation of kink pairs. This model has been adapted to rubies to include Cr-induced hardening for basal slip.

## 2. Experimental procedure

The rubies were provided by R.S.A. Le Rubis (Jarrie, France) under the form of cylinders 1.5–3.5 cm in diameter, grown by the Verneuil technique with four different $Cr^{3+}$ concentrations whose values were determined by PIXE (particle induced X-ray emission) at the CNA (Seville-Spain) and the C2RMF-CNRS (Paris-France). The Cr contents of the four types of rubies are shown in Table 1. Together with the concentration of Cr, the content of other elements, mainly Ti and Ni, has been determined by making use of two softwares (AXIL and Gupix-Win). As shown in Table 1, the results lie within the uncertainty of the PIXE technique. Impurities other than Cr are present in relatively small concentrations. Composition measurements in different points of the surface of the samples indicated relatively good homogeneity.

The crystals were oriented by Laue back reflection X-ray diffraction. They were cut in the form of parallelepipeds of 2 mm × 2 mm × 4 mm. The compression direction is at 57◦ from the [000 1] axis of sapphire, toward the [0 1 1¯ 0] direction (orientation "aa" in Castaing et al.2 ) contained in the (2 1¯ 1¯ 0) plane parallel to two lateral faces. The orientation of the compression load favours the 1/3[1¯ 1¯ 2 0] and 1/3[1¯ 2 1¯ 0] basal slip with a Schmid factor of about 0.4 and it prevents rhombohedral twinning from being activated.2 Further structural information can be found in Refs [8,18,20,21]. In order to permit optical observations of the slip traces left at the surface after straining, the specimens were given a final polish with a diamond paste less than 3 J..m in grain size. The compression tests were performed in air at temperatures between 900 and 1500 ◦C at a constant cross-head speed of 5 J..m/mn, corresponding to an initial strain rate of $\dot{\varepsilon}$ =2.0 × 10−5 s−1 in an Instron machine, model 1185. The load was transmitted to the specimens via SiC rods. Slip lines were analyzed by optical microscopy (Leica DMRE). Foils suitable for transmission electron microscopy (TEM) observations of the dislocation structure were pre- pared by mechanical polishing followed by ion thinning. The TEM observations were performed in a JEOL 200CX electron microscope (Laboratoire d'Etude des Microstructures, CNRS-ONERA, France).

Table 1
Result of the analysis by PIXE techniques of the concentration of impurities in the rubies used in this work

| | Cr concentrations and other impurities (mol ppm) | | | | | | |
|---|---|---|---|---|---|---|---|
| | Cr | Average Cr | Ti | Ni | Cu | Ga | Others |
| **Ruby I** | | | | | | | |
| Axil | 62 ± 13 | 60 | | | | | |
| Gupix-Win | 57 ± 15 | | 5 | 4 | | 4 | 4 (Co) |
| **Ruby II** | | | | | | | |
| Axil | 731 ± 15 | 725 | | | | | |
| Gupix-Win | 718 ± 86 | | 9 | 4 | | 4 | 8 (Co) |
| **Ruby III** | | | | | | | |
| Axil | 3903 ± 38 | 3940 | | | 8 | | |
| Gupix-Win | 3979 ± 187 | | 26 | 9 | 7 | | 5 (Zn) |
| **Ruby IV** | | | | | | | |
| Axil | 9853 ± 168 | 9540 | 47 | | | | |
| Gupix-Win | 9231 ± 1200 | | 45 | 9 | | | 11 (V) |

## 3. Results

Consistent with the crystalline orientation of the specimens and with deformation being achieved by basal slip, Fig. 1 shows sharp slip lines at approximately 30∘ to the compression direction on the $(2\bar{1}\bar{1}0)$ faces (Fig. 1a). Fainter steps, perpendicular to the compression axis, can be observed in the orthogonal faces (Fig. 1b). According to previous works,[2] this suggests that basal slip can be attributed to the activation of the two families of dislocations with $1/3[\bar{1}\bar{1}20]$ and $1/3[\bar{1}2\bar{1}0]$ Burgers vectors in equal numbers. No rhombohedral twinning was observed.

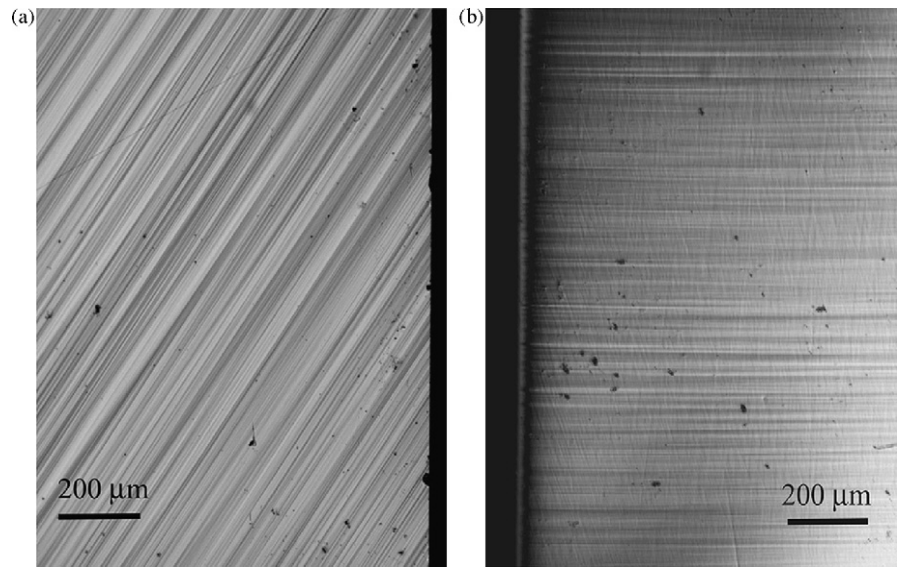

Fig. 1. Optical micrographs of the slip lines in the lateral faces of a sample of ruby containing 9540 mol ppm of Cr and deformed by basal slip at $T = 1000 \circ$ C, with plastic strain $\varepsilon = 2.5\%$ and stress $\sigma = 415$ MPa: (a) the $(2\bar{1}\bar{1}0)$ face and (b) the orthogonal face. The compression direction is vertical, parallel to the edges of the sample faces visible in the micrographs.

The stress–strain curves for the different Cr contents and three temperatures are shown in Fig. 2. Except for the highest temper- ature and Cr content, they generally exhibit a strong drop after the yield point defining an upper and a lower yield point. We take the critical resolved shear stress (CRSS) from the flow stress at the lower yield point. The flow stress increase with Cr content is the most pronounced for the high Cr concentrations. In the two low Cr-doped crystals, i.e. 60 and 725 mol ppm, the flow stresses are very close, actually within the margin of error of our experiments (±1–5 MPa). They are not significantly higher than for undoped sapphire.[2,7] For all Cr concentrations and temperatures, we did not observed any serrated flow that is typical of dynamic strain ageing due to mobile dislocations interacting with diffusing impurities.

As observed in Fig. 2, the Cr content influences the upper yield stress and the flow stress for further deformation, however, not the decrement of engineering stress $!).\sigma$ beyond the upper yield point which varies from $!).\sigma$ = 170 ± 30 MPa at $T$ = 1000 °C to $!).\sigma \approx 0$ at $T$ = 1500 °C, independent of Cr content. This indicates that the multiplication of dislocations and the adjustment of their mean velocity to the imposed strain rate $\dot{\varepsilon}$ are not influenced by the Cr concentration. reduced with increasing temperature. Fig. 2 shows in addition that the work-hardening rate ($!).\sigma/!).\varepsilon$) is little dependent on Cr The region of zero work hardening after the lower point is content. It remains in the range $\mu/70$–$\mu/100$ ($\mu$ = 156 GPa, elastic shear modulus), and compares well to the work-hardening rate found for undoped sapphire deformed along a single $1/3[1\bar{1}\bar{2}0]$ direction in the basal plane.[22] In consequence, the presence of impurities of $Cr^{3+}$ does not seem to affect the work hardening mechanism.[22] The values of the CRSS for all the compression tests are given in Table 2 for the various Cr compositions and temperatures tested (the Schmid factor is 0.4). They are in addition plotted in Fig. 3 as a function of temperature, between 900 and 1500 °C, together with measurements achieved by various authors[2,4,7] for basal slip of undoped sapphire between 600 and 1800 °C. There is a clear increase of the CRSS with Cr addition. Actually, the temperature dependence of the yield stress for sapphire and the rubies are much the same in that the stress increment from one Cr concentration to the next is fairly independent of test temperature. This is more clearly visible when the CRSS is plotted with a linear scale as shown in Fig. 4 where the CRSS of ruby with a Cr concentration of 9540 ppm is about 30 MPa above that of undoped sapphire.

Table 2
Values of the CRSS (MPa) for the lower yield point obtained from the mechanical tests for various Cr compositions and different temperatures of compression tests

| Temperature (°C) | $Cr^{3+}$ concentrations (mol ppm) | | | |
|---|---|---|---|---|
| | 60 | 725 | 3940 | 9540 |
| 900 | | 201 | | |
| 1000 | 122 | 119 | 147 | 162 |
| 1100 | 76 | 74 | 77 / 78 | 96 |
| 1150 | | 54 / 56 | | 77 |
| 1175 | | | | 77 |
| 1200 | 45 / 46 | 44 | 51 / 52 | 66 / 75 |
| 1250 | | | 41 | 66 |
| 1300 | 26 | 26 / 27 | 37 / 37 | 60 |
| 1350 | 21 | 22 | 32 / 33 | 56 |
| 1400 | | | 30 / 31 | 44 |
| 1450 | 14 | 17 | 25 | |
| 1500 | 12 | 15 / 15 | 24 / 25 | 31 / 39 / 41 |

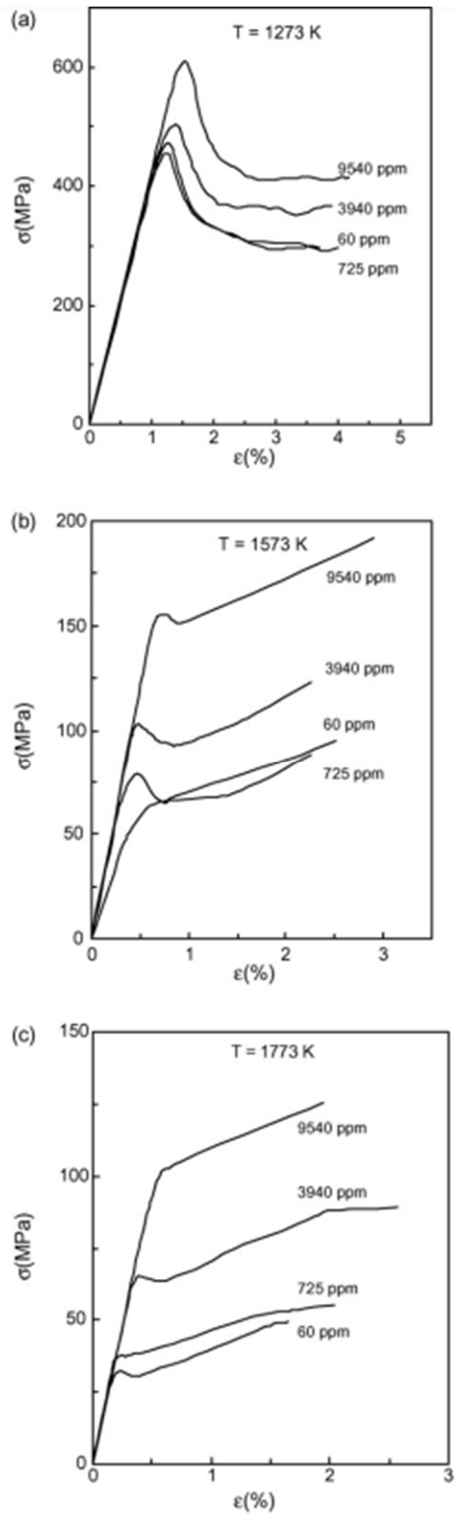

Fig. 2 Engineering stress–strain curves of sapphire doped with different Cr concentrations and deformed by basal slip at different temperatures. In each case, the Cr concentration is indicated: (a) $T = 1000\,°C$; (b) $T = 1300\,°C$ and (c) $T = 1500\,°C$.

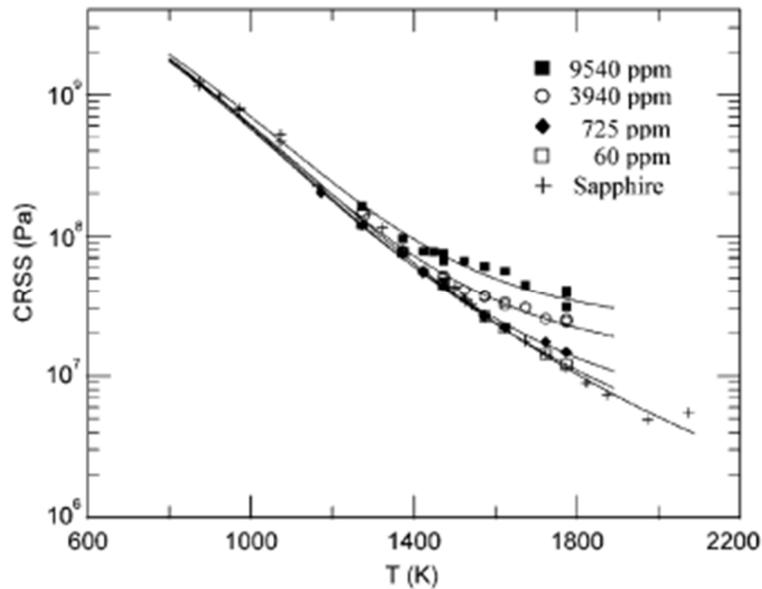

Fig. 3. Plot of log (CRSS) vs. temperature for basal slip in rubies for the different concentrations of Cr. The experimental values obtained by various authors[2,4,7] for basal slip of undoped sapphire have been included. The logarithmic scale makes that the hardening caused by the Cr is hardly visible at low temperatures. The solid curves correspond to the fits from the kink pair model to the experimental values of the CRSS.

The present results agree qualitatively with those published by Pletka et al.[13] who concluded that the Cr solution hardening rate is independent of temperature for Cr concentrations between 100 and 9400 mol ppm between 1300 and 1500 °C. The experimental points obtained by Pletka et al.[13] for a concentration of $Cr^{+3}$ of 9400 mol ppm are included in Fig. 4. It is noted though that for all the Cr concentrations, the CRSS values obtained by Pletka et al.[13] are slightly superior to those obtained in our work. The origin of this difference is twofold, as Pletka et al. have used a higher strain rate ($2.6 \times 10^{-5}$ s$^{-1}$) and their samples, grown by the Verneuil and the Czochralski techniques, contained a higher concentration of other impurities (Si, S, Cl, Ti and Fe) besides Cr.

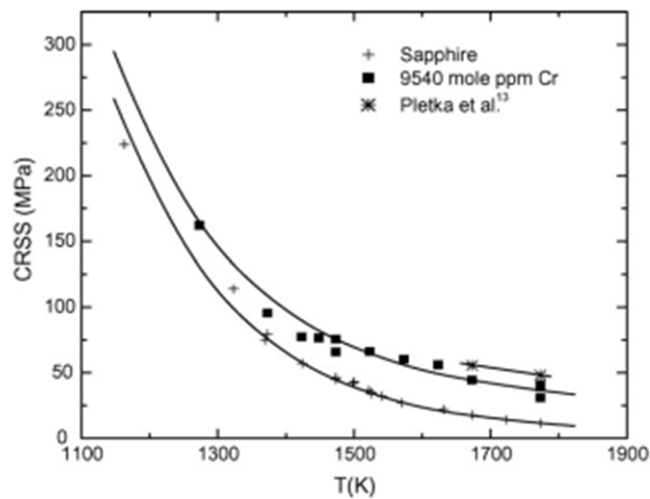

Fig. 4. Experimental values of CRSS vs. temperature for basal slip in sapphire and in rubies containing 9540 mol ppm of Cr and in sapphire. The experimental values obtained by Pletka et al.[13] in ruby containing 9400 mol ppm of Cr are shown.

TEM observations of deformed rubies have been performed in order to assess the part taken by the Peierls mechanism in con- trolling dislocation motion. This role is exemplified in Fig. 5, taken in a sample deformed at 1000 ◦C to 2% (9540 mol ppm of Cr), by the high density of dislocations with long straight segments oriented along (1⁻ 1⁻ 2 0) and  10 1⁻ 0) directions. Other TEM observations of similar dislocation structures have been performed; they will be published in a future publication. These dislocation structures are consistent with previous TEM observations of dislocations after deformation below 800 ◦C3  and with the suggested Peierls mechanisms for the deformation of undoped sapphire.3–6

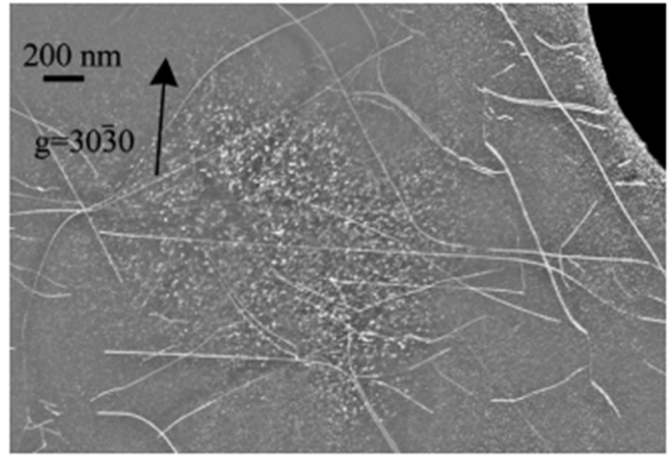

Fig. 5. TEM micrograph of a dislocation structure in ruby containing 9540 mol ppm of Cr and deformed at 1000 ◦ C to 2%. The foil was cut parallel to the (000 1) slip plane.

## 4. Discussion

In metallic solid solutions, solution hardening increases with decreasing temperature below room temperature23 and is approximately independent of temperature at higher tempera- tures. This so-called plateau region is therefore analogous to the temperature insensitivity of Cr solution hardening in ruby between 900 and 1500 ◦C. Mechanisms used to explain solution hardening of metallic systems in the plateau region may therefore be applicable to solution hardening in rubies. This together with the present TEM observations of segmented dis- locations suggest to make use of a model of dislocation glide controlled by the nucleation and propagation of kink pairs (Peierls mechanism) in order to model the plastic properties of ruby in the ranges of temperatures and Cr contents investigated.

### 4.1. Deformation of sapphire and ruby

Rubies can be plastically deformed at temperatures as low as 900 ◦C. The specific orientation employed in the present study prevents rhombohedral twinning from being activated and favours basal slip,2 the single deformation mechanism. The stress strain curves for ruby with 60 ppm Cr (Fig. 2) are identical to those for undoped sapphire; solid solution hardening becomes noticeable only for high chromium content (725 ppm and above). Solution hardening by Cr is moderate. Even the samples with largest Cr concentration, show limited solution hardening (Figs. 2 and 3). The structure of dislocations is dominated by segments along crystallographic directions (Fig. 5) typical of a Peierls mechanism; it does not reveal any feature that would have resulted from strong interactions between dis- locations and Cr impurities.

Athermal Cr solution hardening is observed essentially because plastic deformation can be achieved only at high temperature (above 0.6 times the melting temperature) in sapphire2–5 which is brittle below 900 ◦C. This explains the impossibility to observe in sapphire thermally-activated Cr hardening below 0.2–0.3 times the melting temperature, contrary to observations in metals.23

Although the analysis of work-hardening is beyond the scope of the present investigation, one can make the following remarks. The fact that two slip directions are simultaneously activated in the basal plane does not result in a work-hardening rate larger than in the case of sapphire deformed in single basal slip.3,22 In this case, hardening has been ascribed to the impediment of $1/3[\bar{1}\ \bar{1}\ 2\ 0]$ dislocation slip by prismatic loops, themselves originating from dipole constriction.22 In this context, doping with Cr does not seem to substantially affect the formation of loops and the interactions between these and basal dislocations.

*4.2. Chromium-related obstacles to dislocation glide*

We discuss the possible mechanisms of Cr hardening of sap- phire that a previous study had concluded to be dominated by size effects.13 The physical properties of ruby have been exten- sively explored for a number of reasons, in particular for its remarkable red colour24 which makes it attractive not only for jewels but also for its ability at measuring hydrostatic stresses through the shift of light emission lines.25 The X-ray absorption fine structure (XAFS) in the vicinity of Cr K edge shows that when Cr (ionic radius 0.0615 nm) substitutes for Al (ionic radius

0.0535 nm) in sapphire, the oxygen atoms forming the first shell move apart to accommodate the larger ionic radius (Cr–O dis- tance = 0.197 nm, as for $Cr_2O_3$, compared to Al–O = 0.186 nm), while the distance between a Cr atom and Al atoms belong- ing to the next shell is very close to the Al–Al bond length in sapphire. Atoms further than 0.25 nm apart are almost undis- placed. In other words, atom relaxation around Cr is very local, almost entirely relaxed by the first shell of oxygen atoms. In so far as plasticity is concerned, elastic interactions between Cr atoms and edge dislocations are cancelled beyond the limited volume subject to Cr-induced lattice distortion, in such a way that hardening due to size effect is expected to be substantially less than for an elastic relaxation following a $r^{-1}$ behaviour.23

Another contribution to solution hardening arises from the dif- ference in elastic modulii due to substitutional Cr.23 Roughly speaking, it is for an impurity located at one or two interatomic distances from the dislocation that the maximum elastic energies for the volume interaction and for the elastic misfit interaction ($\mu Cr_2O_3$ = 120 GPa) become comparable. Assuming that dislo- cation pinning is efficient only at such small distances, it cannot be concluded on the hardening mechanism based on classical elastic model that is clearly inadequate at atomic distances.

It is nevertheless interesting to consider the role possibly played by short distance interactions with kinks moving in their Peierls potential. In this case, one would expect substantial changes in the activation energy for kink diffusion along the dislocation line. The use of the kink pair model (see following section) does not bring any evidence that Cr atoms at the core of the dislocations have such an influence on their glide.

The comparison with experimental data is made with the models of Fleicher14 and of Labusch.15 Fleischer's model,14 using the differences in size and shear modulus between solvent and solutes, predicts a dependence of hardening with the atomic concentration of solute c according to the equation !).σ = Kc1/2 while Labusch's,15 using a different statistical treatment, yields a slightly different expression, i.e. !).σ = Kc2/3 . In the next section, such equations are used although the mechanical data can hardly allow checking the validity of the various mechanisms.23

4.3. A kink-pair mechanism for dislocation glide

A kink pair model has been developed by Hirth and Lothe19 to explain dislocation glide controlled by a Peierls mechanism. In this well-known model, dislocation motion takes place by jumps of a segment from a minimum of the lattice potential to the next. The model has been subsequently adapted by Mitchell et al.16,17 to include various situations, such as the glide of dissociated dislocations and the effect of impurities.

In the case of basal slip in ruby and sapphire, the dissociation of the basal dislocations in two 1/3 01 1⁻ 0) partial dislocations generates a stacking-fault with a very high energy,27 so that basal dislocations had rather be regarded as perfect dislocations with an extended core.8 In this case, the kink pair model refers to its version for undissociated dislocations. According to Mitchell et al.,17 the model designed for sapphire postulates that in a high Peierls stress materials the kinks are abrupt with a square shape, the propagation of which along the dislocation line is determined by a secondary Peierls barrier.

In the *long-segment* limit, a kink pair expands until each kink annihilates with kinks of the opposite sign originating from a neighbouring kink source along the line. Following Hirth and Lothe,19 Mitchell et al.17 have derived the strain rate in the case of an undissociated dislocation as

$$\dot{\varepsilon} = \rho a v \frac{\sigma b^2 h^2}{\alpha^{1/2} kT} \exp\left(-\frac{Q_D + F_k}{kT}\right) \exp\left(\frac{(\sigma b^3 h^3 \mu)^{1/2}}{(8\pi)^{1/2} kT}\right) \quad (1)$$

where $\rho$ is the mobile dislocation density, $b$ the Burgers vector of the perfect dislocation, $\sigma$ the resolved shear stress for the glide of dislocations, $h$ the height of the kink (periodicity of the Peierls barriers in the glide plane), $a$ the periodicity along the dislocation line, $v$ the attempt frequency (typically the Debye frequency), $Q_D$ the activation energy for a kink to overcome its secondary Peierls barrier, $F_k$ is the free energy of a single kink on a perfect dislocation, $k$ is the Boltzman constant, $T$ is the absolute temperature, and $\alpha$ a factor given by

$$\alpha = \frac{1}{2}\left\{1 + \left[1 + \left(\frac{2\sigma\mu b^3 h^3}{\pi k^2 T^2}\right)^{1/2}\right]^{1/2}\right\} \quad (2)$$

approximately equal to unity at high temperatures and greater than unity for high stress and low temperature.

In the *short-segment* limit, the dislocation portion on which the kink pair nucleates is sufficiently short for the pair to expand to the full segment length. In this case, the strain rate is given by

$$\dot{\varepsilon} = \rho L \upsilon \frac{\sigma b^2 h^2}{\alpha k T} \exp\left(-\frac{Q_D + 2F_k}{kT}\right) \exp\left(\frac{(\sigma b^3 h^3 \mu)^{1/2}}{(2\pi)^{1/2} kT}\right) \quad (3)$$

where $L$ is the length of the glide dislocation segments arbitrarily taken as $\rho^{-1/2}$.

In order to incorporate the athermal hardening from Cr impu- rities in the model, we proceed by analogy with the theories developed for solution hardening of metallic systems in the plateau regions and we simply replace in the above mentioned Eqs. (1)–(3) the shear stress $\sigma$ by $\sigma l$, where

$$\sigma l = \sigma - \Delta\sigma = \sigma - Kc^n \quad (4)$$

and $n = 1/2$ or $n = 2/3$ depending upon whether we make use of the Fleischer14 or the Labusch15 model, respectively. Eq. (4) assumes that the driving force to nucleate and move kinks is not the applied stress $\sigma$ because a part, $!).\sigma$, of this stress is used to overcome an average elastic field due to solute atoms in con- centration $c$. The presence of the $Cr^{3+}$ solute ions in sapphire is assumed to have no other influence on the kink pair mechanism. We now check the fit between the experimental data and the predictions of the model to evaluate these assumptions. We take $\dot{\varepsilon}$ = 2.0 × 10−5 s−1, $\upsilon$ = 1013 s−1, $b$ = 0.475 nm(Burgers vector of the perfect dislocation), $h$ = $a$ = 0.275 nm (oxygen–oxygen distance in the basal plane). For undoped sap- phire data, we showed that it is not possible to achieve a satisfactory fit over the whole temperature range using a constant dislocation density $\rho$ in Eqs. (1) and (3).28 Indeed, it is not physically acceptable to take the same dislocation density in samples deformed at $T$ = 527 ◦C, with the applied stresses of the order of 103 MPa, and in samples deformed at $T$ = 1727 ◦C, as the applied stresses differ by two orders of magnitude. Instead we assume that at the yield point the dislocation density is proportional to the square of the resolved shear stress following:

$$\rho = \left(\frac{\sigma}{C\mu b}\right)^2 \quad (5)$$

where $C$ is a constant. This expression has been previously employed by Pletka et al. in their investigation of the relation ship between the density of dislocations and the stress applied on deformed sapphire beyond the lower yield point. In expression (5) too, $\sigma l$ is substituted for $\sigma$ since it is $\sigma l$ that is responsible for dislocation multiplication. Attempts at fitting $!).\sigma = Kc^n$ with $n$ = 1/2 and $n$ = 2/3 provide the best results with the latter value (Table 3), in agreement with the conclusions of Pletka et al.13

Least square fits between the experimental CRSS data (Figs. 3 and 4) and the theoretical predictions of Eqs. (1) and (3) have been optimized numerically considering both the *long-segment* and the *short-segment* limits. As expected, several combinations of the set of parameters in Eqs. (1)–(5) give fits of comparable good qualities but the values of the adjustable parameters ($\mu$, $C$, $K$, $\rho$, $Q_D + F_k$, $Q_D + 2F_k$) can vary greatly. In the *short-segment* approach (Eq. (3)), the values of the optimized parameters are not acceptable because they are too far from the data for sapphire and rubies and/or have no reasonable physical justification, as already found noticed for undoped sapphire.28 The *long-segment* approach (Eq. (1)) turns out to be also the most appropriate to describe the temperature dependence of the CRSS in ruby. The best values of the adjustment parameters are shown in Table 3. With the value of the elastic shear modulus, $\mu$ = 156 GPa, the *long-segment* limit fit (fit I, Table 3) yields values of $C$ (Eq. (5)) comprised between 0.054 and 0.039, associated to somewhat high values of dislocation densities. It is noted that Pletka et al.22 obtained larger values, i.e. $C$ = 0.34 at $T$ = 1400 °C down to $C$ = 0.18 at $T$ = 1620 °C. With a value for $C$ fixed in such a way that dislocation densities in ruby compare with those in undoped sapphire,22 fit II (Table 3) is also rather satisfactory implying, however, a value of $\mu$ slightly inferior to the litera- ture values29 (135 GPa at 1500 °C) for sapphire. In both fits (I and II in Table 3), the parameter $K$ is in good agreement with $K$ = 0.93 GPa obtained by Pletka et al.13 for ruby at $T$ = 1500 °C, with $\dot{\varepsilon}$ = 2.6 × 10−5 s−1 and $n$ = 2/3.

Table 3
Values from the obtained parameters of the adjustment of the experimental data of the CRSS of the sapphire and rubies to the kink pair model; I and II correspond to different adjustment conditions in the *long-segment* limit (see the text)

| Parameters | Long-segment limit | | | | | | | | | |
|---|---|---|---|---|---|---|---|---|---|---|
| | I | | | | | II | | | | |
| $c$ (ppm) | 0 | 60 | 725 | 3940 | 9540 | 0 | 60 | 725 | 3940 | 9540 |
| $\mu$ (GPa) | 156$^a$ | | | | | 115 | 110 | 110 | 110 | 125 |
| $C$ | 0.054 | 0.054 | 0.040 | 0.039 | 0.039 | 0.108$^*$ | | | | |
| $K$ (GPa) | – | 0.623 | 0.623 | 0.623 | 0.643 | – | 0.480 | 0.480 | 0.480 | 0.480 |
| $\rho$ (m$^{-2}$) | | | | | | | | | | |
| $T$ = 1500 °C | 2.9 × 10$^{13}$ | 2.7 × 10$^{13}$ | 3.3 × 10$^{13}$ | 4.1 × 10$^{13}$ | 9.1 × 10$^{13}$ | 2.0 × 10$^{12}$ | 2.0 × 10$^{12}$ | 2.0 × 10$^{12}$ | 2.8 × 10$^{12}$ | 4.9 × 10$^{12}$ |
| $T$ = 1000 °C | 1.3 × 10$^{15}$ | 9.2 × 10$^{14}$ | 1.4 × 10$^{15}$ | 2.1 × 10$^{15}$ | 2.1 × 10$^{15}$ | 2.3 × 10$^{14}$ | 2.3 × 10$^{14}$ | 2.0 × 10$^{14}$ | 2.9 × 10$^{14}$ | 3.1 × 10$^{14}$ |
| $Q_D + F_k$ (eV) | 3.6 | 3.6 | 3.6 | 3.6 | 3.7 | 3.4 | 3.4 | 3.4 | 3.4 | 3.5 |

$^a$ Fixed value to make the fit.

The activation energy $Q_D + F_k$ (Eq. (1)) is comprised between 3.6 and 3.7 eV for fit I and between 3.4 and 3.5 eV for fit II. These values can be considered as physically acceptable, in accordance with the literature16–18 that indicates that values of $F_k$ and $Q_D$ of the order of 1 eV are reasonable. The fact that the activation energy in ruby is independent of Cr con- centration, including sapphire,28 is consistent with our initial assumption that the presence of Cr3+ is at the origin of the aver- age stress field and has no influence on the kink pair mechanism (Eqs. (4) and (1)). In fits I and II (Table 3) as well as in many other cases not shown here, the CRSS calculated from the model is always close to the corresponding experimental value. The curves for the fits shown in Fig. 3 were obtained in the *long-segment* approxima- tion (fit II in Table 3) that gives physically meaningful parameter values.

## 5. Conclusions

The yield stress of ruby (a-Al2 O3 doped with Cr3+ ) deformed by basal slip between 900 and 1500 °C does not depend signifi- cantly on Cr content. Dislocation dynamics does not seem to be affected by the presence of Cr which only influences the flow stress. The higher the Cr concentration, the stronger the ruby. Solid solution hardening is approximately constant in the 900–1500 °C range. The athermal nature of solid solution hardening can be explained via elastic interactions between dis- locations and Cr atoms in the a-Al2 O3 lattice.

The work-hardening rate does not depend substantially on Cr content either, which in the interpretation provided by Pletka et al.22 suggests that these impurities affect neither the formation of dipoles and loops nor their interactions with glide dislocations.

TEM observations of ruby samples are consistent with dislocation motion controlled by a Peierls mechanism. The experimental values of the CRSS have been fitted on the basis of the kink pair model. The good quality of the fits between the experimental data and a Peierls model adapted to the present samples strongly suggests that the $Cr^{3+}$ ions do not take part to the kink pair mechanism, at least directly; these ions would contribute only to an average field due to elastic properties of the impurities in the lattice.

**Acknowledgement**

The authors acknowledge the financial support of the Min- istry of Science and Technology (Government of Spain) through the project MAT2003-04199-CO2-02.